\documentclass[aps,prd,reprint,floatfix,superscriptaddress,nofootinbib,longbibliography]{revtex4-2}

\usepackage{amsmath,amssymb,bm,mathtools}
\usepackage{booktabs,array}
\usepackage{graphicx}
\usepackage{xcolor}
\usepackage{silence}
\usepackage{hyperref}
\usepackage{microtype}
\usepackage{placeins}

\definecolor{coolred}{RGB}{164,38,44}
\hypersetup{
colorlinks=true,
linkcolor=coolred,
citecolor=coolred,
urlcolor=coolred,
pdftitle={Coherent and Stochastic Axion Dark Matter from Thermal Relaxation},
pdfauthor={Khan},
pdfsubject={Thermal axion relaxation coherent attenuation stochastic production and collisionless transport},
pdfkeywords={axion dark matter, thermal alignment, fluctuation dissipation, phase space, free streaming}
}
\graphicspath{{figures/}}

\newcommand{\dd}{\mathrm{d}}
\newcommand{\Mp}{M_{\mathrm P}}
\newcommand{\cI}{\mathcal I}

\newcommand{\vk}{\bm{k}}

\newcommand{\vtheta}{\vartheta}
\newcommand{\avg}[1]{\left\langle #1\right\rangle}
\newcommand{\Th}{T_h}
\newcommand{\OmDM}{\Omega_{\mathrm{DM}}}

\begin{document}
\makeatletter
\def\NAT@cmprs{\z@}
\makeatother

\title{Coherent and Stochastic Axion Dark Matter from Thermal Relaxation}

\author{Shahid Hussain Gurmani}
\email{shahidgurmani07@gmail.com}
\affiliation{School of Computer Information Engineering, Shanxi Technology and Business University, Taiyuan, China}

\author{Arzu Cilli}
\email{acilli@yildiz.edu.tr}
\affiliation{Yildiz Technical University, Department of physics, Davutpasa St., 34220, Esenler Istanbul Türkiye}

\author{Phongpichit~Channuie}
\email{phongpichit.ch@mail.wu.ac.th (Corresponding Author)}
\affiliation{School of Science \& College of Graduate Studies, Walailak University, Nakhon Si Thammarat, 80160, Thailand}

\author{Ahmadjon Abdujabbarov}
\email{ahmadjonab@gmail.com}
\affiliation{School of Physics, Harbin Institute of Technology, Harbin 150001, People's Republic of China}

\author{Farruh~Atamurotov}
\email{atamurotov@yahoo.com}
\affiliation{Kimyo International University in Tashkent, Shota Rustaveli str. 156, Tashkent 100121, Uzbekistan}

\author{Ertan G\"{u}dekli}
\email{gudekli@istanbul.edu.tr}
\affiliation{Department of Physics, Faculty of Science, Istanbul University, 34134, Istanbul, Turkey}

\begin{abstract}
An axion driven toward a thermal minimum need not remain a coherent condensate. We derive the coupled attenuation of the field mean and production of stochastic momentum modes in an expanding plasma. A single momentum dependent optical depth partitions every mode between coherent power and Bose occupation, placing standard misalignment, partial relaxation, and stochastic production in one dynamical relation. Finite spatial and temporal response suppresses high momentum production, so the stochastic relic has lower mean momentum and a shorter free streaming length than an equilibrium population at fixed mass and bath temperature. A weakly coupled realization produces the dark matter abundance, reduces inherited coherent isocurvature, and confines production correlations to scales removed by collisionless propagation. Thermal relaxation therefore determines the composition, momentum distribution, and transported structure of axion dark matter.
\end{abstract}

\maketitle

\section{Introduction}

A thermal minimum determines the force on an axion field while leaving the relic's statistical composition unresolved. Coherent misalignment resides in the one point function, whereas stochastic production resides in the connected two point function. Standard misalignment follows the coherent component into nonrelativistic density~\cite{Preskill:1982cy,Abbott:1982af}. A dissipative plasma evolves both field moments, and the abundance integral combines them. Equal abundances can consequently correspond to different coherent fractions, momentum distributions, and spatial power.

Thermal curvature and trapping can displace a scalar toward a temperature dependent minimum~\cite{Batell:2021aja,Batell:2022dpx}. Temporary explicit breaking and thermal phase offsets extend this dynamics to axion initial conditions and primordial isocurvature~\cite{Brandenberger:2025axm}~\cite{Batell:2026phase}. Gauge transport adds dissipation to coherent axion motion~\cite{Choi:2022fbu,Papageorgiou:2022chm}, while an open thermal environment also populates stochastic modes~\cite{Cao:2022rjr}. These ingredients determine separate parts of the evolution, but they do not specify the momentum resolved relation among coherent attenuation, stochastic occupation, and the spatial distribution retained after decoupling.

Four dynamical operations act in sequence and on different field moments. An integrated restoring strength determines whether the angle moves far enough before the thermal curvature disappears. The least damped local eigenvalue determines coherent survival. A momentum dependent optical depth determines the stochastic occupation generated by the bath. The collisionless transfer integral determines which spatial correlations reach gravitational scales. A peak comparison of the thermal mass with the Hubble rate addresses none of these operations in full, while the total abundance removes their statistical distinction.

We derive a modewise relation in which the fractional Bose occupation and retained coherent power add to unity. The relation connects coherent misalignment, partial relaxation, and stochastic production before momentum integration. Finite spatial and temporal response makes the optical depth decrease with momentum. This monotonicity orders every increasing kinematic moment and implies that the generated population has no larger mean momentum or free streaming length than Bose equilibrium at fixed mass and bath temperature. The relation therefore determines both relic composition and the velocity support generated by thermal relaxation.

A scalar multiplet generates a transient periodic curvature, while a weak hidden gauge bath generates dissipation and stochastic forcing. Their static and transport correlators probe different limits of the thermal spectral function. The realized parameter set produces the dark matter abundance mainly through stochastic modes, strongly reduces the inherited coherent isocurvature component, and generates a momentum spectrum concentrated below Bose equilibrium. Its production correlations occupy microscopic scales that collisionless propagation removes before gravitational structure develops. The paper determines the finite time alignment dynamics, derives the modewise composition and kinematic ordering, and propagates the resulting distribution into abundance, velocity support, density power, and isocurvature.
\section{Thermal realization}
\label{secThermal}

Write the periodic field as $\theta=a/f_a$, where $f_a$ is the decay constant. The Peccei Quinn mechanism promotes the strong interaction angle to a dynamical field~\cite{Peccei:1977hh,Peccei:1977ur}, whose pseudoscalar excitation is the axion~\cite{Weinberg:1977ma,Wilczek:1977pj}. This field description applies to the QCD axion and broader axionlike particles~\cite{Sikivie:2006ni,Marsh:2015xka}, with low energy couplings determined by their ultraviolet realization~\cite{Hook:2018dlk,DiLuzio:2020wdo}. The transient thermal contribution is
\begin{equation}
V_T(\theta,T)=f_a^2m_T^2(T)
\left[1-\cos(\theta+\beta)\right],
\label{eqThermalPotential}
\end{equation}
where $\beta$ is the angular separation between the thermal and late vacuum minima. The shifted angle $\vtheta=\theta+\beta$ measures displacement from the thermal minimum.

A scalar multiplet $\chi_A$ with $N_\chi$ real components generates the curvature through
\begin{equation}
\Delta V_T=
\frac{\lambda_1f_a^2}{M_\star^2}
(\chi_A\chi_A)^2
\left[1-\cos(\theta+\beta)\right].
\label{eqScalarOperator}
\end{equation}
For $y=m_\chi/T$, define
\begin{equation}
F_2(y)=\frac{6}{\pi^2}
\int_0^\infty\dd x\,
\frac{x^2}{
\sqrt{x^2+y^2}
\left[\exp\!\left(\sqrt{x^2+y^2}\right)-1\right]}.
\label{eqFtwo}
\end{equation}
Each scalar component has variance $T^2F_2(y)/12$. Gaussian contraction gives
\begin{equation}
\avg{(\chi_A\chi_A)^2}_T=
\frac{N_\chi(N_\chi+2)}{144}T^4F_2^2(y),
\label{eqRadialMoment}
\end{equation}
and therefore
\begin{equation}
m_T^2(T)=
\frac{\lambda_1N_\chi(N_\chi+2)}{144}
\frac{T^4}{M_\star^2}F_2^2(m_\chi/T).
\label{eqThermalMass}
\end{equation}
The relativistic regime has $F_2(0)=1$ and $m_T^2\propto T^4$. Once $T$ approaches $m_\chi$, the scalar occupation decreases and the restoring curvature vanishes continuously. The Higgs portal
\begin{equation}
\Delta\mathcal L_{H\chi}=-\frac{\lambda_{H\chi}}{2}
|H|^2\chi_A\chi_A
\label{eqHiggsPortal}
\end{equation}
maintains scalar equilibrium, with relativistic rate
\begin{equation}
\Gamma_\chi\simeq\frac{\lambda_{H\chi}^2}{64\pi}T.
\label{eqScalarRate}
\end{equation}
The benchmark satisfies the equilibrium condition underlying Eq.~\eqref{eqRadialMoment}~\cite{Batell:2021aja,Batell:2022dpx}. Similar medium induced mass shifts can link coherent scalar evolution with particle freeze out~\cite{Ferrante:2026coherent}.

Equation~\eqref{eqScalarOperator} contributes to the renormalized periodic potential at the matching scale. Planck suppressed breaking and higher dimension terms can shift a late axion minimum~\cite{Holman:1992us,Kamionkowski:1992mf}, while their renormalized coefficients depend on the ultraviolet theory~\cite{Barr:1992qq,Dobrescu:1996jp}. At the subtraction scale $\mu_R=m_\chi$, the finite scalar threshold generated by Eq.~\eqref{eqScalarOperator} is
\begin{equation}
\delta m_0^2=
\frac{\lambda_1N_\chi(N_\chi+2)}{M_\star^2}
\left(\frac{m_\chi^2}{16\pi^2}\right)^2.
\label{eqVacuumCurvature}
\end{equation}
The benchmark gives $\delta m_0^2/m_a^2=5.54\times10^{-13}$. The associated displacement of the late minimum is bounded by this ratio times $|\sin\beta|$. When the late state is identified with a QCD axion, the threshold lies below the neutron electric dipole constraint~\cite{Dine:2022qgf,Abel:2020pzs}. A vacuum term of comparable size can lift wall degeneracy without setting the late mass~\cite{Zhang:2023rnp}. The appendix derives Eq.~\eqref{eqVacuumCurvature}.

The dissipative response arises from a hidden Yang Mills bath,
\begin{equation}
\mathcal L_h\supset
\frac{\alpha_h}{8\pi}\frac{a}{f_h}
G^a_{h\mu\nu}\widetilde G_h^{a\mu\nu}.
\label{eqGaugePortal}
\end{equation}
While the gauge plasma is active, its low frequency response is
\begin{equation}
\Upsilon_h(\Th)=
\kappa_\gamma(N_g\alpha_h)^5
\frac{\Th^3}{f_h^2}\,f_{\rm act}(\Th),
\label{eqFriction}
\end{equation}
with
\begin{equation}
f_{\rm act}(\Th)=\frac{1}{2}
\left[1+\tanh\left(
\frac{\ln(\Th/T_{h,*})}{\Delta_h}
\right)\right].
\label{eqActiveFraction}
\end{equation}
A hidden Higgs condensate gives the gauge bosons mass near $T_{h,*}$ and suppresses the topological response while the coupling remains perturbative. Smooth hyperbolic tangent, error function, and compact cubic transition profiles produce indistinguishable integrated yields at the accuracy relevant to the abundance calculation. The appendix compares these profiles.

The scalar expectation value and the gauge topological density enter through different limits of thermal correlation functions. Equation~\eqref{eqThermalMass} is a static angular self energy, whereas Eq.~\eqref{eqFriction} probes the low frequency spectral density paired with stochastic forcing. Topological transport can remain active in a deconfined plasma even when the equilibrium susceptibility is suppressed~\cite{Choi:2022fbu,Papageorgiou:2022chm}. Near confinement, the susceptibility follows nonperturbative gauge dynamics~\cite{Borsanyi:2016ksw,Bonati:2015vqz}. This separation isolates the restoring force from the transport coefficient while allowing both to arise from different spectral limits of one microscopic sector. A gluodynamic scalar sector provides one such effective realization~\cite{Pirzada:2026uak}.

The expansion rate includes the Standard Model plasma, the scalar multiplet, and the hidden bath,
\begin{equation}
H^2(T)=\frac{\pi^2T^4}{90\Mp^2}
\left[g_{*\rho}^{\rm SM}(T)+g_{*\rho}^{\chi}(T)
+g_{*\rho}^{h}(T)\right].
\label{eqHubble}
\end{equation}
The Standard Model functions follow the thermal equation of state across the electroweak and QCD crossovers~\cite{Laine:2015kra,Saikawa:2018rcs}. Visible entropy conservation gives
\begin{equation}
\frac{\dd\ln T}{\dd N}=-
\left[1+\frac{1}{3}
\frac{\dd\ln g_{*s}^{\rm SM}}{\dd\ln T}\right]^{-1},
\qquad N=\ln(a/a_i).
\label{eqEntropy}
\end{equation}
The hidden sector obeys
\begin{equation}
g_{*s}^{h}(\Th)\Th^3a^3={\rm constant}.
\label{eqHiddenEntropy}
\end{equation}
The active plasma has $g_{*s}^{h}=52$. After the hidden gauge and radial modes become massive, the residual fermions have $g_{*s}^{h}=42$, which increases $\Th a$ by a factor $1.074$. More than $0.9999995$ of the stochastic yield is produced before this transition. The decoupled axion momenta then redshift independently, while the residual hidden radiation gives $\Delta N_{\rm eff}=5.98\times10^{-3}$.

The benchmark describes a stable photophobic axionlike particle with an independent late mass. Its two photon width is
\begin{equation}
\Gamma_{a\gamma\gamma}=\frac{g_{a\gamma}^2m_a^3}{64\pi}.
\label{eqPhotonWidth}
\end{equation}
A lifetime longer than the age of the Universe requires
\begin{equation}
g_{a\gamma}<3.66\times10^{-15}\ {\rm GeV}^{-1}
\left(\frac{2.83\times10^2\ {\rm keV}}{m_a}\right)^{3/2}.
\label{eqStabilityBound}
\end{equation}
The visible coupling is taken below Eq.~\eqref{eqStabilityBound}, and every hidden two body threshold lies above $m_a$, placing the benchmark outside the interaction ranges usually targeted by axion searches~\cite{Irastorza:2018dyq,ChadhaDay:2021szb}. Its mass and decay constant also lie outside the standard QCD band~\cite{DiLuzio:2020wdo}. The QCD susceptibility that controls conventional axion cosmology~\cite{Petreczky:2016vrs,Berkowitz:2015aua} therefore does not determine the independent late mass used here.

\begin{table}[t]
\caption{Benchmark parameters and dimensionless validity tests.}
\label{tabBenchmark}
\centering
\begin{ruledtabular}
\begin{tabular}{lc}
Quantity & Value\\
\colrule
$N_\chi$ & $16$\\
$\lambda_1$ & $6.00$\\
$\lambda_{H\chi}$ & $2.00\times10^{-3}$\\
$m_\chi$ & $1.00\times10^4\ {\rm GeV}$\\
$M_\star$ & $1.25\times10^{17}\ {\rm GeV}$\\
$f_a$ & $2.50\times10^8\ {\rm GeV}$\\
$f_h$ & $2.50\times10^8\ {\rm GeV}$\\
$\xi_{h,i}$ & $2.50\times10^{-1}$\\
$m_a$ & $2.83\times10^2\ {\rm keV}$\\
$\min(\Gamma_\chi/H)$ & $6.65\times10^1$\\
$\max(\alpha_h)$ & $3.77\times10^{-2}$\\
$T_i/M_\star$ & $1.60\times10^{-9}$\\
$\delta m_0^2/m_a^2$ & $5.54\times10^{-13}$\\
$\max(\rho_a/\rho_{\rm tot})$ & $1.76\times10^{-7}$\\
$\Delta N_{\rm eff}$ & $5.98\times10^{-3}$
\end{tabular}
\end{ruledtabular}
\end{table}

The ratios in Table~\ref{tabBenchmark} establish the validity hierarchy. The higher dimension expansion is governed by $T_i/M_\star$, scalar equilibrium by $\Gamma_\chi/H$, and perturbative transport by $\alpha_h$. The axion energy fraction bounds its contribution to the expansion rate and bath spectral density. Each approximation is therefore tied to a separate dimensionless ratio rather than to the abundance result.
\section{Alignment dynamics}
\label{secAlignment}

The homogeneous angle obeys
\begin{equation}
\vtheta''+q(N)\vtheta'
+\mu^2(N)\sin\vtheta=0,
\label{eqMeanEquation}
\end{equation}
with
\begin{equation}
q(N)=3+\frac{\dd\ln H}{\dd N}+\frac{\Upsilon_h}{H},
\qquad
\mu^2(N)=\frac{m_T^2}{H^2}.
\label{eqQMu}
\end{equation}
The coefficient $q$ controls angular mobility, whereas $\mu^2$ measures restoring curvature relative to expansion. Their local competition is resolved by the harmonic amplitude decay rate
\begin{equation}
\Gamma_A(N)=\frac{1}{2}
\left[q(N)-{\rm Re}\sqrt{q^2(N)-4\mu^2(N)}\right],
\label{eqAmplitudeRate}
\end{equation}
and by its integrated value
\begin{equation}
\mathcal A(N)=\int_0^N\Gamma_A(s)\,\dd s.
\label{eqAmplitudeAction}
\end{equation}
For $q^2<4\mu^2$, the field is underdamped and $\Gamma_A=q/2$. In the strongly overdamped regime, $\Gamma_A\simeq\mu^2/q$, so additional friction reduces angular motion. The critical curve $q=2\mu$ maximizes $\Gamma_A/\mu$. A large peak value of $m_T/H$ cannot distinguish these regimes.

A complementary nonlinear exclusion criterion follows directly from Eq.~\eqref{eqMeanEquation}. For $q(N)\geq q_0>0$ and initial rest, define
\begin{equation}
\cI(N)=\frac{1}{q_0}\int_0^N\mu^2(s)\,\dd s.
\label{eqAction}
\end{equation}
The cosine equation satisfies
\begin{equation}
\frac{\displaystyle
\sup_{0\leq s\leq N}|\vtheta(s)-\vtheta_i|}
{|\vtheta_i|}
\leq\frac{\cI(N)}{1-\cI(N)},
\qquad \cI(N)<1.
\label{eqNonlinearBound}
\end{equation}
A target residual $|\vtheta_f|\leq\epsilon|\vtheta_i|$ therefore requires
\begin{equation}
\cI\geq\frac{1-\epsilon}{2-\epsilon}.
\label{eqNecessaryAction}
\end{equation}
The appendix derives this bound from the retarded velocity equation. Equation~\eqref{eqNecessaryAction} excludes insufficient integrated curvature. Equations~\eqref{eqAmplitudeRate} and~\eqref{eqAmplitudeAction} then identify whether the available curvature produces oscillatory damping, critical relaxation, or overdamped arrest. The two integrals answer different questions. The quantity $\cI$ bounds how far the nonlinear field can move, whereas $\mathcal A$ integrates the least damped local mode and measures coherent survival. Large restoring strength can coexist with weak attenuation when the trajectory remains deeply overdamped.

The standard anharmonic misalignment relation follows the coherent trajectory but contains no bath occupation~\cite{Dine:1982ah,Turner:1985si}. The pair $(\cI,\mathcal A)$ remains useful when early curvature arises from hidden monopoles~\cite{Kawasaki:2017xwt}, when angular momentum stores the initial condition~\cite{Co:2019wyp}, or when a transient barrier delays release~\cite{DiLuzio:2021gos,DiLuzio:2024sor}. Topological forcing and nonstandard thermal histories modify the angular trajectory and the interval over which relaxation acts~\cite{Banerjee:2024bua}~\cite{Arias:2022qjt,Ramazanov:2022imq}. The restoring integral measures the available displacement, whereas the amplitude integral measures the survival of the least damped mode, so mechanisms with similar peak masses need not leave equal coherent fractions.

\begin{figure*}[t]
\centering
\includegraphics[width=0.487\textwidth]{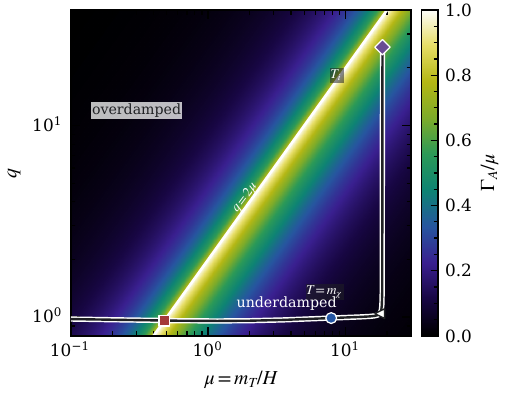}
\hfill
\includegraphics[width=0.487\textwidth]{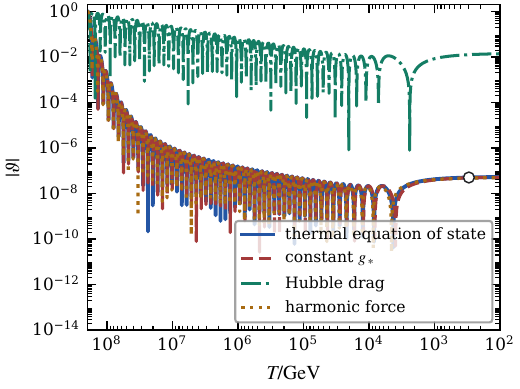}
\caption{Damping regimes and nonlinear alignment. The left panel shows the local decay efficiency $\Gamma_A/\mu$ in the $(\mu,q)$ plane. The white curve is critical damping and the black trajectory follows the benchmark from the initial temperature through scalar decoupling. The right panel compares the nonlinear thermal history with constant relativistic degrees of freedom, Hubble drag, and harmonic curvature. The open marker identifies late mass activation.}
\label{figAlignment}
\end{figure*}

Figure~\ref{figAlignment} makes the relaxation mechanism explicit. The benchmark begins in the underdamped region and accumulates most of its amplitude decay while $\Gamma_A/\mu$ remains appreciable. The scalar population then disappears, $\mu$ falls, and the trajectory crosses the critical ridge into the overdamped region. Beyond that crossing, $\Gamma_A\simeq\mu^2/q$ and further attenuation rapidly ends. The critical ridge therefore identifies the part of the thermal history that converts restoring curvature into amplitude decay most efficiently.

The benchmark has $\mathcal A=19.1$ and $\cI=3.15\times10^3$, reaching $|\vtheta_f|=5.33\times10^{-8}$ from $\vtheta_i=1$. Hubble drag alone leaves $|\vtheta_f|=1.38\times10^{-2}$. Temperature dependent thermodynamics changes the final amplitude by about three percent relative to constant relativistic degrees of freedom. Across the abundance matching scan, $\Omega_{a,{\rm coh}}/\Omega_{\rm DM}<4.11\times10^{-6}$ for $\mathcal A\geq20$. The threshold near twenty marks sustained attenuation that is insensitive to the oscillation phase at scalar decoupling.

Inflationary and reheating dynamics determine the long wavelength correlations and initial displacement entering Eq.~\eqref{eqMeanEquation}~\cite{Ijaz:2026ear,Pirzada:2026jml}. Dilaton coupled reheating and prolonged inflation provide distinct prethermal states~\cite{Pirzada:2026sle,Khan:2026doo}. Once the plasma forms, the subsequent attenuation is governed by the thermal response derived above.
\section{Phase space production}
\label{secPhaseSpace}

The hidden bath has finite temporal and spatial correlation scales. The normalized temporal correlator is
\begin{equation}
K_t(\Delta t)=\frac{1}{2\tau_c}
\exp\!\left(-\frac{|\Delta t|}{\tau_c}\right),
\qquad
S_t(\omega)=\frac{1}{1+\omega^2\tau_c^2}.
\label{eqTemporalResponse}
\end{equation}
The nonlocal response and symmetric noise kernel follow the influence functional description of a system coupled to a thermal environment~\cite{Feynman:1963fq,Caldeira:1983rp}. Colored noise retains the finite correlation time~\cite{Hu:1992dv}, while the fluctuation dissipation relation determines the noise normalization from the dissipative response~\cite{Callen:1951vq,Kubo:1966nj}.

Two positive finite range profiles parameterize the spatial response,
\begin{equation}
S_s^{\rm G}(k)=\exp(-k^2\ell^2),
\qquad
S_s^{\rm E}(k)=\frac{1}{(1+k^2\ell^2)^2}.
\label{eqSpatialResponse}
\end{equation}
The second expression is the transform of a three dimensional exponential correlator. Both approach unity at small momentum and differ in their ultraviolet decrease. The microscopic scales are written as
\begin{equation}
\ell^{-1}=c_m g_h^2\Th,
\qquad
\tau_c^{-1}=c_\omega g_h^4\Th.
\label{eqBathScales}
\end{equation}
The dimensionless coefficients retain the microscopic uncertainty in the hidden topological correlator. The yield spread between the two profiles quantifies its effect on the relic calculation.

The local expansion is controlled by
\begin{equation}
\epsilon_c=\tau_c\max
\left(H,m_T,\Upsilon_h,\Gamma_{\rm rel}\right),
\qquad
\Gamma_{\rm rel}=\frac{\omega_k^2}{3H+\Upsilon_h}.
\label{eqMarkovControl}
\end{equation}
The benchmark has $\max\epsilon_c=2.12\times10^{-7}$, placing the evolution in the local transport regime used for thermally damped scalar fields~\cite{Banerjee:2025damp}. Finite correlation time corrections are governed by the expansion parameter that also appears in nonlocal cosmological fluctuation dynamics~\cite{Gangopadhyay:2026nonlocal}. The appendix compares the exponential response with its Lorentzian reduction for the present dark matter distribution.

For $x=p/\Th$, Eq.~\eqref{eqHiddenEntropy} leaves $x$ constant during the active relativistic interval. Local plasma scattering defines a quasiparticle sector
\begin{equation}
x\geq x_{\rm q}\equiv
\max_{N_i\leq N\leq N_f}\frac{H(N)}{\Th(N)}.
\label{eqQuasiparticleDomain}
\end{equation}
Modes below $x_{\rm q}$ remain in the long wavelength coherent field and enter the isocurvature transfer instead of the local gain and loss equation. The horizon therefore separates a field theoretic long wavelength sector from a kinetic subhorizon sector. Extending the local collision term below this boundary conflates inherited isocurvature with plasma generated occupation. Within the weak coupling quasiparticle sector, the occupation obeys
\begin{equation}
\frac{\partial n_x}{\partial N}=\gamma_x(N)
\left[n_{\rm B}(x)-n_x\right],
\label{eqKinetic}
\end{equation}
where $n_{\rm B}(x)=(e^x-1)^{-1}$ and
\begin{equation}
\gamma_x(N)=\frac{\Upsilon_h}{H}
S_s\!\left(\frac{x}{c_mg_h^2}\right)
S_t\!\left(\frac{x}{c_\omega g_h^4}\right).
\label{eqModeRate}
\end{equation}
The integrated optical depth is
\begin{equation}
\tau_x=\int_{N_i}^{N_f}\gamma_x(N)\,\dd N.
\label{eqOpticalDepth}
\end{equation}
For an initially unoccupied stochastic sector,
\begin{equation}
n_x=\left(1-e^{-\tau_x}\right)n_{\rm B}(x).
\label{eqOccupation}
\end{equation}
The coherent amplitude of the corresponding linear mode satisfies $|A_x/A_{x,i}|^2=e^{-\tau_x}$. Hence
\begin{equation}
\boxed{
\frac{n_x}{n_{\rm B}(x)}+
\left|\frac{A_x}{A_{x,i}}\right|^2=1
}.
\label{eqComplementarity}
\end{equation}
This identity is evaluated before any momentum integral. At $\tau_x=0$, the mode remains coherent and reproduces standard misalignment. Finite $\tau_x$ gives a mixed coherent and stochastic state. At large $\tau_x$, the stochastic occupation approaches $n_{\rm B}$. In correlation function language, damping contracts the one point function while noise increases the connected two point function along a common optical depth. The normalized unit sum constrains state composition, while physical energy exchange follows the bath response and cosmological redshift. A relic density integral cannot distinguish these regimes, whereas Eq.~\eqref{eqComplementarity} determines their composition at every momentum.

The finite response also orders the kinematics of the produced population. Define $R_x=1-e^{-\tau_x}$ and the Bose number average
\begin{equation}
\avg{F}_{\rm B}=
\frac{\displaystyle\int\dd\ln x\,x^3n_{\rm B}(x)F(x)}
{\displaystyle\int\dd\ln x\,x^3n_{\rm B}(x)}.
\label{eqBoseAverage}
\end{equation}
Both response factors in Eq.~\eqref{eqModeRate} decrease with $x$ throughout the domain in Eq.~\eqref{eqQuasiparticleDomain}, so $R_x$ is nonincreasing. For any increasing kinematic function $F(x)$,
\begin{equation}
\avg{F}_{\rm prod}-\avg{F}_{\rm B}
=
\frac{{\rm Cov}_{\rm B}(F,R_x)}{\avg{R_x}_{\rm B}}
\leq0.
\label{eqResponseOrdering}
\end{equation}
The inequality follows from the opposite ordering of $F$ and $R_x$. Taking $F=x$ bounds the mean momentum. Taking $F=v(x,a)$ bounds the mean speed at every scale factor and therefore the free streaming length. A finite range bath thus generates a stochastic population that is no hotter than Bose equilibrium at fixed mass and bath temperature. The ordering follows analytically from response monotonicity and precedes any numerical discretization of the distribution.

After tracing over the bath, each linear mode is a displaced Gaussian state. Its coherent amplitude resides in the field mean, while $n_x$ determines the covariance and the bosonic entropy
\begin{equation}
s_x=(1+n_x)\ln(1+n_x)-n_x\ln n_x.
\label{eqModeEntropy}
\end{equation}
Using $\partial n_x/\partial\tau_x=n_{\rm B}-n_x$ gives
\begin{equation}
\frac{\partial s_x}{\partial\tau_x}=
(n_{\rm B}-n_x)\ln\!\left(\frac{1+n_x}{n_x}\right)\geq0
\label{eqEntropyIncrease}
\end{equation}
for $0<n_x\leq n_{\rm B}$. The field mean and covariance are independent components of a displaced Gaussian state. Equation~\eqref{eqComplementarity} constrains their joint evolution through one optical depth, while Eq.~\eqref{eqEntropyIncrease} identifies the direction of relaxation. The optical depth therefore links mean attenuation with covariance growth at each momentum. Expansion ends the interaction at finite optical depth, preserving a momentum dependent displacement and covariance that cannot be represented by one effective temperature. This coupled mean and covariance evolution is the Gaussian open system structure familiar from stochastic scalar dynamics~\cite{Gleiser:1993ea,Miyamoto:2013aka}, thermal damping, and warm inflation~\cite{Yokoyama:2004pf,Berera:1995ie}. Nonlocal formulations retain the bath correlation function explicitly~\cite{Berera:2008ar}.

The perturbative evaluation applies for $\alpha_h\ll1$, $\epsilon_c\ll1$, and $\rho_a/\rho_{\rm tot}\ll1$. The maximum axion energy fraction is $1.76\times10^{-7}$, so the generated population does not modify the bath spectral density or expansion rate at the retained order. The long wavelength coherent sector and the subhorizon quasiparticle sector are therefore evolved by different equations and joined through their contributions to the late density and isocurvature. A nonperturbative hidden spectral function can replace the perturbative response near confinement. Its spectral shape changes $\tau_x$, while the gain and loss structure preserves the mode composition relation.

\begin{figure*}[t]
\centering
\includegraphics[width=0.487\textwidth]{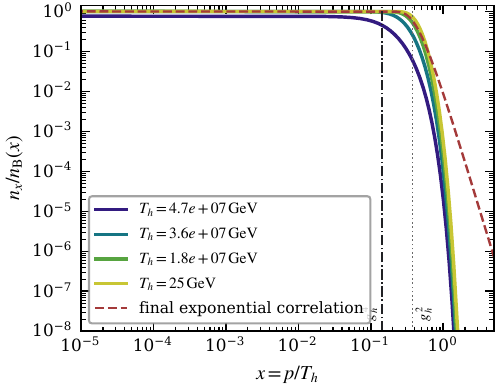}
\hfill
\includegraphics[width=0.487\textwidth]{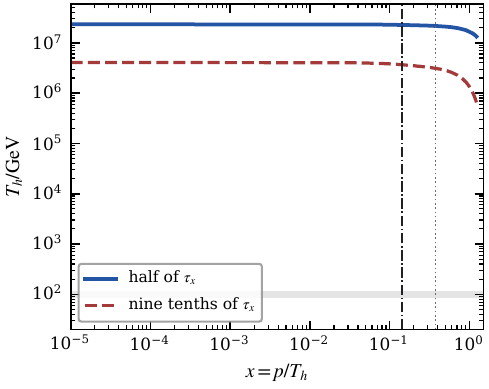}
\caption{Momentum dependent stochastic production. The left panel shows the occupation relative to the Bose distribution at successive hidden temperatures. The dashed curve gives the final exponential spatial response. Vertical lines mark the temporal and spatial response momenta. The right panel gives the hidden temperatures at which each mode accumulates one half and nine tenths of its final optical depth. The shaded band marks the hidden Higgs transition.}
\label{figPhaseSpace}
\end{figure*}

Figure~\ref{figPhaseSpace} resolves how the distribution is assembled. Soft modes accumulate optical depth early because both response factors remain near unity. Modes near the temporal and spatial response momenta acquire their optical depth later and remain underoccupied when the hidden Higgs transition removes the bath. The two knees therefore arise from distinct correlation scales. This momentum dependence controls the number and velocity moments independently.
\section{Abundance and transport}
\label{secTransport}

The stochastic yield is
\begin{equation}
Y_a=\frac{45}{4\pi^4g_{*s}(T_f)}
\left(\frac{\Th}{T}\right)^3
\int_0^\infty\dd\ln x\,x^3n_x.
\label{eqYield}
\end{equation}
The late mass that reproduces the observed abundance is
\begin{equation}
m_a^{\rm DM}=\frac{\rho_{\rm DM}/s_0}{Y_a}.
\label{eqMatchingMass}
\end{equation}
The Gaussian and exponential spatial responses give $Y_a=1.59\times10^{-6}$ and $1.50\times10^{-6}$, corresponding to $m_a^{\rm DM}=2.75\times10^2\ {\rm keV}$ and $2.91\times10^2\ {\rm keV}$. Their smooth profiles span a six percent yield interval. The dominant contribution comes from the soft momentum band populated before the hidden transition.

The residual homogeneous mode contributes
\begin{equation}
Y_{\rm coh}=\frac{m_af_a^2\bar\vtheta^2(T_m)}{2s(T_m)},
\label{eqCoherentYield}
\end{equation}
where $T_m$ is the late mass activation temperature. The benchmark gives
\begin{equation}
f_{\rm coh}\equiv\frac{\Omega_{\rm coh}}{\OmDM}
=1.21\times10^{-5}.
\label{eqCoherentFraction}
\end{equation}
The dark matter abundance is therefore dominated by the stochastic population, unlike conventional misalignment where a coherent angle sets the relic density~\cite{Wantz:2009it}. Nonzero momentum axions can also arise from resonant temperature modulation or postinflationary quenches~\cite{Pirzada:2026npl,Khan:2026nsz}. In the dissipative regime considered here, Eq.~\eqref{eqComplementarity} links the coherent and stochastic populations through one momentum dependent optical depth.

\begin{figure*}[t]
\centering
\includegraphics[width=0.487\textwidth]{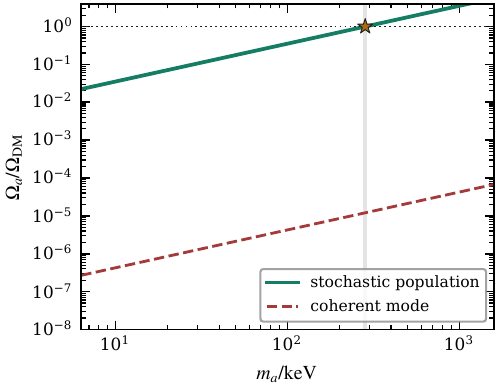}
\hfill
\includegraphics[width=0.487\textwidth]{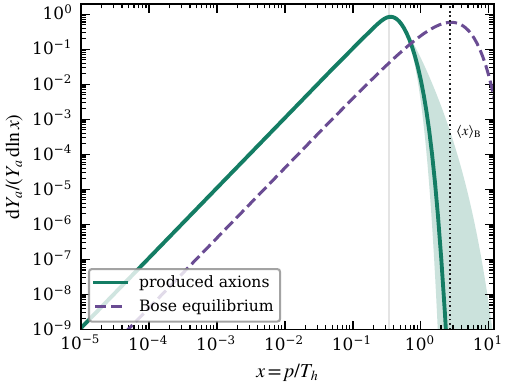}
\caption{Abundance composition and momentum compression. The left panel compares the stochastic abundance with the residual coherent contribution as the late axion mass varies. The shaded band gives the masses obtained from the two smooth spatial responses and the star marks the benchmark. The right panel compares the normalized stochastic yield spectrum with a Bose equilibrium distribution. The finite response concentrates the abundance at soft momentum.}
\label{figAbundance}
\end{figure*}

The right panel of Fig.~\ref{figAbundance} evaluates the kinematic ordering in Eq.~\eqref{eqResponseOrdering}. The mean momentum is $\avg{x}=0.333$ for the Gaussian response and $0.353$ for the exponential response, compared with $\avg{x}_{\rm B}=2.70$ in Bose equilibrium. The ratios are $0.123$ and $0.131$. Matching the initial velocity moment to a Bose relic gives an equivalent thermal mass between $2.17$ and $2.30\ {\rm MeV}$. With $p(a)=xT_{a0}/a$, the calculated spectrum determines the nonrelativistic velocity distribution at any later epoch by horizontal rescaling. The stochastic population therefore has lower momentum than an equilibrium relic with the abundance matching mass.

After decoupling, the comoving free streaming length is
\begin{equation}
\lambda_{\rm fs}=\int_{a_d}^{1}
\frac{\avg{v(a)}}{a^2H(a)}\,\dd a,
\label{eqFreeStreaming}
\end{equation}
where $\avg{v(a)}$ is evaluated from the calculated distribution. The two smooth responses give $\lambda_{\rm fs}=1.96\times10^{-4}\ {\rm Mpc}$ and $2.07\times10^{-4}\ {\rm Mpc}$, corresponding to $k_{\rm fs}$ near $3.1\times10^4\ {\rm Mpc}^{-1}$. The mean speed at equality is about $8\times10^{-8}$ and falls below $2.5\times10^{-11}$ today. High redshift Lyman alpha analyses constrain conventional thermal relic distributions~\cite{Irsic:2023free}. A direct mass translation is inappropriate here because the optical depth produces a subthermal spectrum. The calculated velocity moment and free streaming integral provide the relevant comparison.

Abundance and free streaming probe different moments of $n_x$. The number moment determines $m_a^{\rm DM}$, whereas the momentum moment determines propagation. Their separation is a direct consequence of the finite bath response. The combination of stochastic dominance with a low momentum spectrum distinguishes this production history from coherent misalignment and Bose equilibration. A large coherent isocurvature component or structure suppression above the calculated free streaming scale excludes the benchmark realization.
\section{Density statistics}
\label{secDensity}

A random phase axion field has two independent Gaussian quadratures. Let their variance be $\sigma^2$ and let one quadrature have coherent amplitude $\mu$. The local harmonic density can be written through a noncentral chi square variable $X$ with $\nu$ quadratures and noncentrality $\lambda=\mu^2/\sigma^2$. The normalized density is
\begin{equation}
1+\delta=\frac{X}{\nu+\lambda}.
\label{eqDensityVariable}
\end{equation}
Its variance and normalized skewness are
\begin{equation}
\sigma_\delta^2=\frac{2(\nu+2\lambda)}{(\nu+\lambda)^2},
\qquad
\gamma_1=\frac{8(\nu+3\lambda)}{[2(\nu+2\lambda)]^{3/2}}.
\label{eqDensityMoments}
\end{equation}
The centered random phase limit has unit variance and $\gamma_1=2$. A phase restricted field with one quadrature has $\gamma_1=2\sqrt2$. These local statistics distinguish a random phase population from a residual coherent displacement, but they do not determine the spatial power. The one point probability distribution depends on the number of field quadratures, whereas gravitational structure depends on correlations between separated points.

For quadrature covariance $C_k$ and coherent amplitude $\mu$, Wick contraction gives
\begin{equation}
P_\delta(k)=
\frac{4\mu^2C_k+2\nu\displaystyle
\int\frac{\dd^3q}{(2\pi)^3}C_qC_{|\vk-\bm q|}}
{(\mu^2+\nu\sigma^2)^2}.
\label{eqDensityPower}
\end{equation}
A finite correlation length makes $C_k$ analytic near $k=0$. Consequently
\begin{equation}
P_\delta(k)=P_0+\mathcal O(k^2),
\qquad
\Delta_\delta^2(k)=\frac{k^3P_\delta(k)}{2\pi^2}
\propto k^3.
\label{eqInfraredPower}
\end{equation}
The third power infrared tail is the analytic consequence of finite correlation length. This causal structure appears in finite domain formation and quench dynamics~\cite{Kibble:1976sj,Zurek:1985qw}, including numerical quenches~\cite{Laguna:1996hf} and finite range stochastic correlators~\cite{Durrer:2003ja}.

Particle discreteness adds
\begin{equation}
P_{\rm P}=\frac{1}{\bar n_0},
\qquad
\Delta_{\rm P}^2(k)=\frac{k^3}{2\pi^2\bar n_0}.
\label{eqPoisson}
\end{equation}
The benchmark has $\bar n_0=1.31\times10^{71}\ {\rm Mpc}^{-3}$. The sum of the continuous field and particle contributions remains below $5.55\times10^{-59}$ at $k_{\rm fs}$.

\begin{figure*}[t]
\centering
\includegraphics[width=0.487\textwidth]{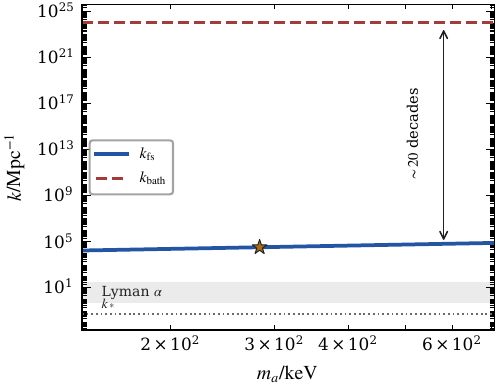}
\hfill
\includegraphics[width=0.487\textwidth]{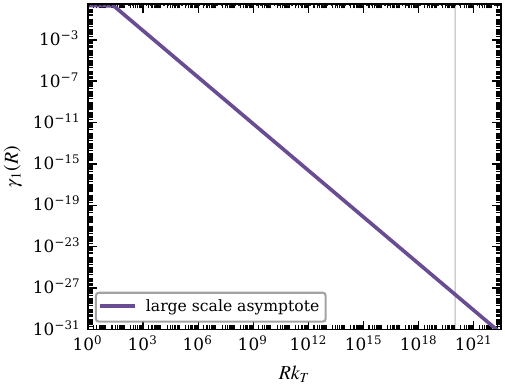}
\caption{Microscopic production and gravitational smoothing. The left panel compares the CMB pivot scale, the high redshift Lyman alpha interval, the free streaming wavenumber, and the bath correlation wavenumber across the abundance matching mass range. The right panel shows the analytic decrease of normalized skewness under spatial averaging. The shaded band marks the free streaming radius.}
\label{figDensity}
\end{figure*}

For a Gaussian window of radius $R$ and finite spectral moments, the variance scales as $R^{-3}$ and the third connected moment scales as $R^{-6}$. Hence
\begin{equation}
\gamma_1(R)\propto R^{-3/2}
\qquad Rk_T\gg1.
\label{eqSkewnessScaling}
\end{equation}
The microscopic bath wavenumber is $1.05\times10^{24}\ {\rm Mpc}^{-1}$, approximately twenty decades above $k_{\rm fs}$. Figure~\ref{figDensity} also places the free streaming cutoff above the CMB pivot and the high redshift Lyman alpha interval~\cite{Planck:2018jri,Irsic:2023free}. The normalized skewness at the free streaming radius is near $2.4\times10^{-28}$. Spatial averaging therefore suppresses the local skewness before gravity can retain the fluctuations. The result separates non Gaussian one point statistics from cosmological density power. A broad local probability distribution does not imply compact structure when its correlation length is microscopic.

This scale hierarchy gives a sharp structure result. Axion miniclusters require order unity angular variations on gravitational scales~\cite{Hogan:1988mp,Kolb:1993zz}, with their mass spectrum inherited from the initial spatial power~\cite{Enander:2017ogx} and its nonlinear evolution~\cite{Vaquero:2018tib}. Postinflationary strings and walls can generate those correlated initial conditions~\cite{Gorghetto:2018myk,Kawasaki:2014sqa}, whose nonlinear spectrum has been resolved in large simulations~\cite{Buschmann:2019icd,Dine:2020pds}. The microscopic correlations produced by thermal relaxation remain far below this gravitational regime, so the benchmark does not generate minicluster seeds despite its non Gaussian local density distribution.

Large scale isocurvature constrains both inherited coherent perturbations and long wavelength modulation of the bath~\cite{Planck:2018jri}. For an inherited coherent mode $S_{\rm coh}$ that is uncorrelated with stochastic production,
\begin{equation}
S_{\rm DM}=f_{\rm coh}S_{\rm coh},
\qquad
P_{S,{\rm DM}}(k)=f_{\rm coh}^2P_{S,{\rm coh}}(k).
\label{eqIsocurvatureTransfer}
\end{equation}
Equation~\eqref{eqCoherentFraction} gives a power transfer factor $1.45\times10^{-10}$.

A hidden radiation entropy mode changes the yield through the temperature ratio $\xi_h=\Th/T$. Define
\begin{equation}
R_\xi\equiv\frac{\partial\ln Y_a}{\partial\ln\xi_h},
\qquad
S_{h\gamma}=3\,\delta\ln\xi_h.
\label{eqBathResponse}
\end{equation}
The total long wavelength perturbation is
\begin{equation}
S_{\rm DM}=f_{\rm coh}S_{\rm coh}
+\frac{R_\xi}{3}S_{h\gamma}.
\label{eqBathIsocurvature}
\end{equation}
The numerical derivative gives $R_\xi=4.02$ and $4.23$ for the two smooth spatial responses. The Planck bound on an uncorrelated cold dark matter entropy mode then requires the root mean square hidden radiation entropy perturbation below $6.5\times10^{-6}$ to $6.8\times10^{-6}$. Adiabatic visible and hidden radiation has $S_{h\gamma}=0$, while a primordial hidden entropy mode is constrained by Eq.~\eqref{eqBathIsocurvature}. Background data independently constrain residual radiation and free streaming effects~\cite{Planck:2018vyg}.

The two terms in Eq.~\eqref{eqBathIsocurvature} apply across inflationary initial states~\cite{Ijaz:2026ear,Pirzada:2026jml}, quenched non Abelian production~\cite{Khan:2026nsz}, and temperature modulated axion production~\cite{Pirzada:2026npl}. Each history enters through either the inherited coherent perturbation or the hidden radiation entropy mode, so the response coefficients preserve their distinct physical origins.
\section{Conclusion}

Thermal relaxation changes the statistical state of axion dark matter. Conventional misalignment is specified by one coherent angle, whereas the relaxed relic requires a field mean and a momentum dependent covariance. The nonlinear restoring integral determines whether the mean can move, the least damped eigenvalue determines its survival, and the optical depth determines covariance growth. Their separation explains why an aligned trajectory and a matched abundance leave the late phase space underdetermined.

Within linear response, the optical depth is the scalar coordinate linking coherent and stochastic components at each momentum. It contracts coherent power and fills Bose occupation in complementary fractions. Monotonic finite response then orders every increasing kinematic moment, yielding a stochastic spectrum colder than Bose equilibrium at fixed mass and bath temperature. In the weakly coupled realization, this spectrum produces the dark matter abundance while the surviving coherent fraction strongly reduces inherited isocurvature. The predicted population is a momentum resolved stochastic relic rather than a rescaled thermal distribution.

Spatial statistics require an additional transfer. The bath correlator determines microscopic density structure, while free streaming determines which correlations remain on gravitational scales. Their separation suppresses density power, particle discreteness, and normalized skewness above the free streaming length, while the yield response quantifies hidden radiation isocurvature. Microscopic local non Gaussianity therefore coexists with a cosmologically smooth, subthermal relic. Thermal relaxation leaves a phase space state whose coherent fraction, stochastic spectrum, velocity support, isocurvature response, and gravitational density power follow from distinct moments of the underlying dynamics.
\appendix

\section{Derivations and numerical validation}
\label{appDerivations}

For one free scalar component, the thermal variance is
\begin{equation}
I_T(m_\chi)=\int\frac{\dd^3p}{(2\pi)^3}
\frac{1}{E_p\left(e^{E_p/T}-1\right)}
=\frac{T^2}{12}F_2(m_\chi/T).
\label{eqThermalTadpole}
\end{equation}
Rotational symmetry in field space gives
\begin{equation}
\avg{(\chi_A\chi_A)^2}_T
=N_\chi(N_\chi+2)I_T^2,
\label{eqThermalWick}
\end{equation}
which reproduces Eqs.~\eqref{eqRadialMoment} and~\eqref{eqThermalMass}. In dimensional regularization, the finite vacuum tadpole is
\begin{equation}
I_0(\mu_R)=\frac{m_\chi^2}{16\pi^2}
\left[\ln\!\left(\frac{m_\chi^2}{\mu_R^2}\right)-1\right].
\label{eqVacuumTadpole}
\end{equation}
Contracting the two scalar pairs in Eq.~\eqref{eqScalarOperator} gives
\begin{equation}
\delta m_0^2(\mu_R)=
\frac{\lambda_1N_\chi(N_\chi+2)}{M_\star^2}I_0^2(\mu_R).
\label{eqVacuumThresholdGeneral}
\end{equation}
At $\mu_R=m_\chi$, Eq.~\eqref{eqVacuumThresholdGeneral} reduces to Eq.~\eqref{eqVacuumCurvature}. Expanding the combined late and threshold potentials around the late minimum gives $|\delta\theta_{\rm min}|\leq(\delta m_0^2/m_a^2)|\sin\beta|$ at leading order.

Hidden entropy density is $s_h=2\pi^2g_{*s}^{h}\Th^3/45$. Adiabatic hidden evolution gives $\dd(s_ha^3)/\dd t=0$ and therefore
\begin{equation}
\Th a\left(g_{*s}^{h}\right)^{1/3}={\rm constant}.
\label{eqHiddenEntropySolution}
\end{equation}
For the benchmark, $g_{*s}^{h}$ changes from $52$ to $42$, so $\Th a$ increases by $(52/42)^{1/3}=1.074$. The cumulative occupation integral reaches its asymptotic value before the hidden transition. The phase space distribution has therefore decoupled before the entropy release.

Set $v=\vtheta'$ in Eq.~\eqref{eqMeanEquation}. The velocity equation is
\begin{equation}
v'+q(N)v=-\mu^2(N)\sin\vtheta.
\label{eqVelocity}
\end{equation}
For initial rest, the integrating factor gives
\begin{equation}
v(N)=-\int_0^N\dd s\,
\exp\!\left[-\int_s^Nq(y)\,\dd y\right]
\mu^2(s)\sin\vtheta(s).
\label{eqVelocityRetarded}
\end{equation}
A second integration yields
\begin{align}
\vtheta(N)-\vtheta_i
={}&-\int_0^N\dd s\,\mu^2(s)
\sin\vtheta(s)K(N,s),
\label{eqVolterra}\\
K(N,s)={}&\int_s^N\dd z\,
\exp\!\left[-\int_s^zq(y)\,\dd y\right].
\label{eqKernelBoundDef}
\end{align}
The condition $q\geq q_0$ gives $0\leq K(N,s)\leq q_0^{-1}$. Define
\begin{equation}
D_N=\sup_{0\leq z\leq N}|\vtheta(z)-\vtheta_i|.
\label{eqDN}
\end{equation}
Using $|\sin x|\leq|x|$ and $|\vtheta(s)|\leq|\vtheta_i|+D_N$ in Eq.~\eqref{eqVolterra} gives
\begin{equation}
D_N\leq\cI(N)(|\vtheta_i|+D_N).
\label{eqBoundIntermediate}
\end{equation}
Rearrangement for $\cI<1$ gives Eq.~\eqref{eqNonlinearBound}. The target condition in Eq.~\eqref{eqNecessaryAction} follows from $|\vtheta_f-\vtheta_i|\geq(1-\epsilon)|\vtheta_i|$.

Linearizing Eq.~\eqref{eqMeanEquation} over a short interval with locally constant $q$ and $\mu$ gives characteristic exponents
\begin{equation}
r_\pm=-\frac{q}{2}\pm\frac{1}{2}\sqrt{q^2-4\mu^2}.
\label{eqLocalRoots}
\end{equation}
The least negative real part gives Eq.~\eqref{eqAmplitudeRate}. When $q^2<4\mu^2$, both roots have real part $-q/2$. When $q^2\gg4\mu^2$, expansion of the square root gives $\Gamma_A=\mu^2/q+\mathcal O(\mu^4/q^3)$. The integrated rate in Eq.~\eqref{eqAmplitudeAction} therefore distinguishes oscillatory damping from slow overdamped drift. Refined quadrature and independent integration reproduce the benchmark value $\mathcal A=19.1$. Along the abundance matching band, the region $\mathcal A\geq20$ has $\Omega_{a,{\rm coh}}/\Omega_{\rm DM}<4.11\times10^{-6}$. The intermediate region retains sensitivity to the oscillation phase at decoupling, whereas larger integrated decay gives phase insensitive attenuation.

A generalized Langevin mode can be written as
\begin{equation}
\ddot\vtheta_k+3H\dot\vtheta_k+\omega_k^2\vtheta_k
+\int_{t_i}^{t}\dd s\,K(t-s)\dot\vtheta_k(s)=\xi_k(t).
\label{eqGeneralLangevin}
\end{equation}
For $K(t)=\Upsilon_h e^{-t/\tau_c}/\tau_c$, introduce
\begin{equation}
\dot z_k=-\frac{z_k}{\tau_c}
+\frac{\Upsilon_h}{\tau_c}\dot\vtheta_k
\label{eqAuxResponse}
\end{equation}
and the colored force
\begin{equation}
\dot\xi_k=-\frac{\xi_k}{\tau_c}
+\frac{\sqrt{2\Upsilon_h\Th}}{\tau_c}\zeta_k,
\label{eqAuxNoise}
\end{equation}
where $\zeta_k$ is normalized white noise after the cosmological volume factor is restored. Eliminating $z_k$ and $\xi_k$ reproduces the finite response and paired noise correlator. Expansion in $\omega\tau_c$ gives the Lorentzian factor in Eq.~\eqref{eqTemporalResponse}.

\begin{figure*}[t]
\centering
\includegraphics[width=0.487\textwidth]{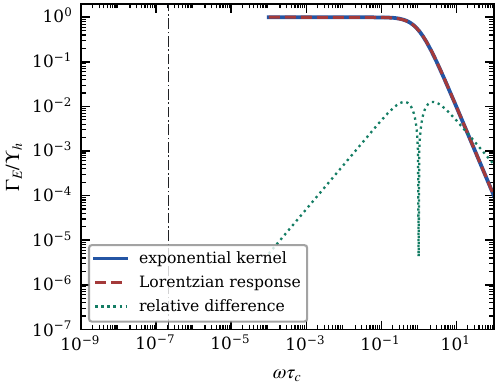}
\hfill
\includegraphics[width=0.487\textwidth]{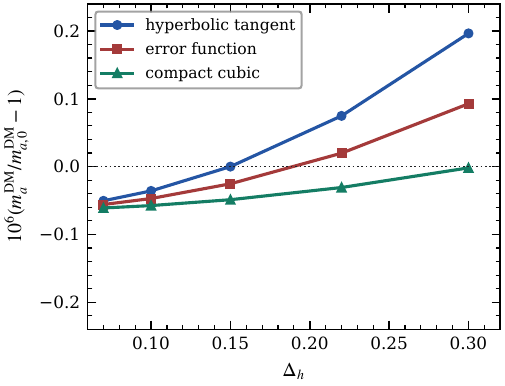}
\caption{Finite response and transition profile tests. The left panel compares the exponential response with the Lorentzian reduction and their relative difference. The vertical line marks the largest benchmark value of $\omega\tau_c$. The right panel gives the abundance matching mass for three smooth hidden Higgs transitions. The benchmark lies within the local response regime and the matching mass is stable under the transition shape.}
\label{figAppendixChecks}
\end{figure*}

In the overdamped quasiparticle regime,
\begin{equation}
\dot n_k=\Gamma_k\left(n_{{\rm B},k}-n_k\right).
\label{eqGainLossTime}
\end{equation}
Changing variables from $t$ to $N$ and inserting the spatial and temporal factors gives Eqs.~\eqref{eqKinetic} and~\eqref{eqModeRate}. Since $x=p/\Th$ is conserved while $a\Th$ is constant, $n_{\rm B}(x)$ has no explicit $N$ dependence. Direct integration gives
\begin{equation}
n_x(N_f)=e^{-\tau_x}n_x(N_i)
+\left(1-e^{-\tau_x}\right)n_{\rm B}(x).
\label{eqGeneralOccupation}
\end{equation}
The homogeneous solution has $A_x=e^{-\tau_x/2}A_x(N_i)$. Setting $n_x(N_i)=0$ gives Eq.~\eqref{eqComplementarity}. Differentiating Eq.~\eqref{eqModeEntropy} with $\partial n_x/\partial\tau_x=n_{\rm B}-n_x$ gives Eq.~\eqref{eqEntropyIncrease}.

Let $g_j$ be independent Gaussian quadratures with variance $\sigma^2$, and let one coherent component have amplitude $\mu$. The variable
\begin{equation}
X=\frac{(\mu+g_1)^2+\sum_{j=2}^{\nu}g_j^2}{\sigma^2}
\label{eqNoncentralVariable}
\end{equation}
has a noncentral chi square distribution with noncentrality $\lambda=\mu^2/\sigma^2$. Rescaling by its mean gives Eq.~\eqref{eqDensityVariable}. The probability density is
\begin{equation}
p(\delta\mid\nu,\lambda)=
(\nu+\lambda)
 f_{\chi'^2_\nu}\!\left[(\nu+\lambda)(1+\delta)\right],
\qquad \delta\geq-1.
\label{eqDensityPdf}
\end{equation}
The cumulants $\kappa_1=\nu+\lambda$, $\kappa_2=2(\nu+2\lambda)$, and $\kappa_3=8(\nu+3\lambda)$ give Eq.~\eqref{eqDensityMoments}. For $\nu=2$ and $\lambda=0$, the local density is exponential with unit variance and skewness two. For $\nu=1$, the centered skewness is $2\sqrt2$.

The density power follows from Wick contraction. For $\vtheta_j=\mu\delta_{j1}+g_j$, the Fourier density fluctuation is
\begin{equation}
\delta\rho_a(\vk)\propto
2\mu g_1(\vk)+
\sum_{j=1}^{\nu}
\int\frac{\dd^3q}{(2\pi)^3}
g_j(\bm q)g_j(\vk-\bm q),
\label{eqDensityFourier}
\end{equation}
with the zero momentum expectation removed. The linear contraction gives $4\mu^2C_k$, the connected quadratic contraction gives the convolution in Eq.~\eqref{eqDensityPower}, and the mixed term vanishes. For Gaussian smoothing with $W_R(k)=\exp(-k^2R^2/2)$,
\begin{equation}
\sigma_R^2\simeq
\frac{P_0}{8\pi^{3/2}R^3}
\label{eqSmoothedVariance}
\end{equation}
when $R$ exceeds the correlation length. The connected third moment scales as $R^{-6}$ when its spectral moment is finite, which gives Eq.~\eqref{eqSkewnessScaling}.

For a large scale coherent perturbation uncorrelated with the stochastic population, write $\rho_{\rm DM}=\rho_{\rm coh}+\rho_{\rm stoch}$ and $\delta\rho_{\rm stoch}=0$ on that scale. Then
\begin{equation}
S_{\rm DM}=\frac{\delta\rho_{\rm DM}}{\rho_{\rm DM}}
=\frac{\rho_{\rm coh}}{\rho_{\rm DM}}S_{\rm coh},
\label{eqIsocurvatureDerivation}
\end{equation}
which gives Eq.~\eqref{eqIsocurvatureTransfer}. For radiation, the relative entropy perturbation satisfies $S_{h\gamma}=3(\delta\Th/\Th-\delta T/T)=3\delta\ln\xi_h$. A stochastic yield perturbation is therefore
\begin{equation}
\delta\ln Y_a=R_\xi\delta\ln\xi_h
=\frac{R_\xi}{3}S_{h\gamma},
\label{eqBathIsocurvatureDerivation}
\end{equation}
which gives Eq.~\eqref{eqBathIsocurvature}. Symmetric derivatives evaluated at successively smaller hidden temperature perturbations reproduce the quoted response coefficients. For an uncorrelated entropy mode, $\beta_{\rm iso}=P_S/(P_{\mathcal R}+P_S)$. With $\beta_{\rm iso}<0.038$ and $P_{\mathcal R}=2.10\times10^{-9}$, the root mean square dark matter entropy perturbation is below $9.11\times10^{-6}$. Equation~\eqref{eqBathIsocurvature} then gives the hidden radiation bounds quoted in the main text.

The finite response approaches the Lorentzian reduction throughout the benchmark domain. The smooth transition profiles in Fig.~\ref{figAppendixChecks} produce coincident abundance matching masses and mean field trajectories at the displayed precision. The axion energy fraction remains small throughout production, preserving the linear response hierarchy.

Independent adaptive integration, refined momentum and angular quadrature, and separate symbolic reconstruction reproduce the abundance, velocity moments, transport scales, and analytic identities. The phase space integral begins at the quasiparticle boundary in Eq.~\eqref{eqQuasiparticleDomain}. Lowering the numerical integration limit below that boundary leaves the number moment and density power normalization unchanged. An order one variation of the boundary leaves the abundance and velocity moments invariant, while the local third moment retains the specified separation between coherent long wavelength modes and stochastic quasiparticles. A finite correlation ensemble independently reproduces the large volume scaling in Eq.~\eqref{eqSkewnessScaling}. These checks are documented with the numerical material accompanying the article.

\FloatBarrier

\bibliography{references}

\end{document}